D.V. Vlasov, L.A. Apresian, V.I. Krystob, T.V. Vlasova

# Nonlinear response and two stable electrical conductivity levels measured in plasticized PVC thin film samples

The electrical conductivity of PVC films prepared with a patented plasticizer of type "A" was measured with high precision automated setup, based on standard ring sell with a voltage range much less than breakdown voltage. Continual voltage-current measurements permit to take into account Debay relaxation process and clearly distinguish specific polymer film conductivity effects, connected with continuous current-stabilization behavior and transitions between two stable (long-living) states with several order magnitude different conductivities. Spontaneous reversible and non-destructive transitions of resistance levels was observed. For 30 mkm polymer films the values of sample resistance was measured equal to: high- $10^6$ Ohm and low - $10^3$ Ohm.

Электропроводность ПВХ пленок, пластифицированных с помощью патентованного модификатора типа «А», исследовалась экспериментально с использованием автоматизированной высокоточной установки и стандартной кольцевой ячейки, при напряженностях поля значительно меньших уровня пробоя. Измерения вольт-амперных характеристик в непрерывном режиме позволило корректно учесть релаксационные дебаевские процессы и выдклить эффекты, связанные с нелинейностью и переходами между двумя состояниями с разными уровнями электропроводности. Полученные нелинейные характеристики отвечают стабилизации тока в состояниях с высоким сопротивлением. Наблюдались спонтанные и неразрушающие переходы между состояниями с высокой и низкой проводимостью: для полимерных пленок толщиной 30 мкм уровни сопротивления составляли $10^3$ Ом и $10^6$ Ом, соответственно.

**Введение**

Электрические явления в полимерных пленок представляет значительный интерес как для выяснения физических механизмов их электропроводности, так и для решения многих прикладных задач.

Обычно в справочниках значения проводимости ПВХ с различными пластификаторами указывается с разбросом несколько порядков и отмечается, что проводимость зависит от многих факторов, в том числе и от приложенного внешнего поля [1,2]. Известно так же, что в определенных условиях в широком классе полимерных пленок может скачкообразно изменяться сопротивление с переходом пленки в высоко-проводящее состояние [2,4]. Поэтому, хотя формально записать закон Ома для полимерной пленки можно, величина электросопротивления не является «хорошей» измеряемой характеристикой, поскольку она зависит как от величины приложенного поля, так и от интервала времени, прошедшего с момента приложения поля (известный процесс запаздывания отклика, называемого Дебаевской релаксацией) . Тем не менее, ввиду особой важности для приложений этой характеристики полимерных материалов, существуют ГОСТированные методы и приборы, например (тераомметр Е6-13 со стандартной кольцевой ячейкой), для измерения отношения сопротивления к току (эффективного удельного поверхностного и объемного сопротивления образца). Перспективная возможность использования пластифицированного ПВХ для создания пленок нанокомпозитов потребовала более тщательного экспериментального анализа и исследования нелинейных процессов и характерных времен релаксации в пластифицированных ПВХ пленках.

**Экспериментальная установка**

С этой целью был собран автоматизированный комплекс измерения сопротивлений в диапазоне от десятков ГОм до единиц кОм с постоянным отслеживанием и накоплением данных измерений в компьютерном файле. В отличие от стандартных измерителей, работающих при фиксированном напряжении источника, измерительный комплекс позволял варьировать в широких пределах прикладываемое напряжение. При этом в описываемых экспериментах всегда использовались поля существенно меньше порога пробоя, значения которых известны из литературы, и проверялись в отдельных измерениях на соответствующей аппаратуре. Диапазон изменений напряжений был на порядок меньше пробойного и составлял от 0 до 60В при толщине пленок 30-50 мкм. Пленки ПВХ изготавливались как со стандартным пластификатором диоктилфталатом (ДОФ ), так и с использованием нового пластификатора (модификатора типа «А» [3]) методом полива на стеклянные плоские подложки из 4% раствора ПВХ с пластификатором в тетрагидрофуране (ТГФ). Соотношение ПВХ и пластификатора составляло 100:80 (вес.%, соответственно ). Для получения отсчетов напряжения и тока образца использовалась стандартная кольцевая ячейка от ГОСТированного прибора Е6-13 с полной заменой измерительной части на автоматизированный комплекс на базе микропроцессора C8051F120 с встроенным 12 разрядным АЦП и программно управляемым предусилителем при быстродействии ядра процессора до 100 МГц. Интервал отбора отсчетов тока и напряжения составлял 1-2 мин и превосходил все характерные времена установления в измерительном комплексе, что позволило , как показали эксперименты, детально отслеживать(запаздывающий) установившийся отклик проводимости образца на приложенное поле.

**Результаты измерений**

В ходе измерений образцы пленок ПВХ со стандартным пластификатором ДОФ показали высокое значение сопротивления образца 8 ГОм или удельное сопротивление $2,4*10^{11}$ Ом м . При этом значения измеряемых величин хорошо укладывается в диапазон соответствующих табличных значений для ПВХ, пластифицированного ДОФ. Описанные ниже аномалии проводимости наблюдались в образцах ПВХ пленок, пластифицированных модификатором типа «А», и не наблюдались в образцах с ДОФ.

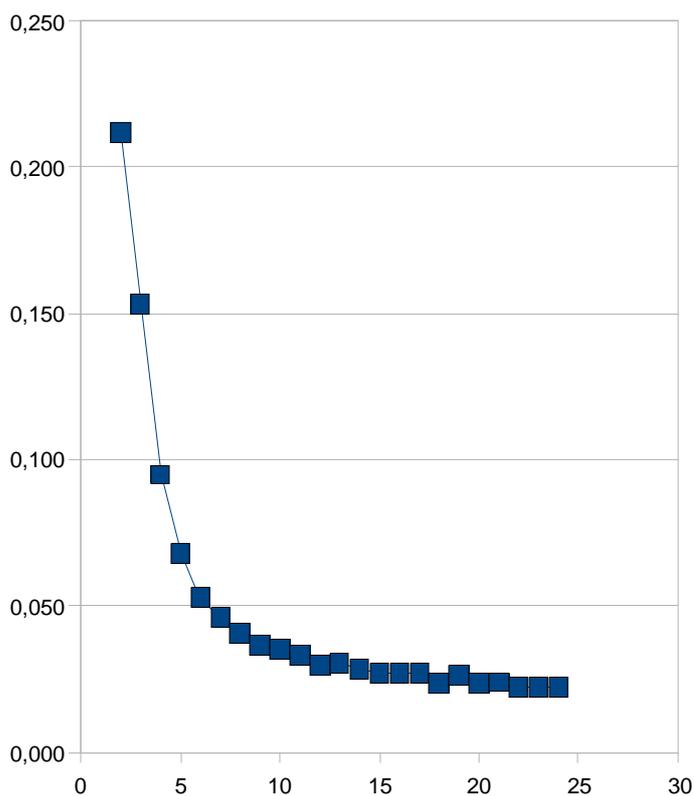

Рис. 1 Релаксация проводимости в течение получасовых измерений

Благодаря непрерывному отслеживанию аппаратным комплексом электросопротивления измеряемого образца удалось установить и разделить несколько эффектов и корректно учесть влияние эффектов релаксации, т.е. запаздывающего отклика полимерных пленок на приложенное поле. На рис.1 показан временной отклик пленки на приложенное напряжение 60В. Особенно сильные изменения наблюдаются у только-что приготовленной пленки, а при определенной «тренировке» в электрических полях (даже существенно ниже порога пробоя) перепад релаксационных изменений проводимости становится существенно меньше.

Второй экспериментальный результат связан с нелинейностью установившегося отклика пленки ПВХ при попытке построения вольт-амперной характеристики. В частности, пленка обнаруживает свойства «стабилизатора тока» - при изменении приложенного напряжения от 5 до 60 В, значения тока через образец практически не изменяются или изменяются, значительно (на порядок) меньше. Такое поведение образца приведено на Рис.2. Можно отметить также гистерезисную зависимость электро-сопротивления при повышении и понижении напряжения, что может быть также обусловлено релаксационными процессами с большими характерными временами. Однако даже длительный процесс измерений, обеспечивающий наблюдаемую релаксацию измеряемых величин, сохраняет стабилизирующий эффект полимерных образцов. Каждая точка на рис. 2 получена усреднением по 10 последовательным точкам (соответственно время измерений каждой точки не менее 10 мин) для усреднения по «релаксационным» дрейфам электрических параметров образца.

.

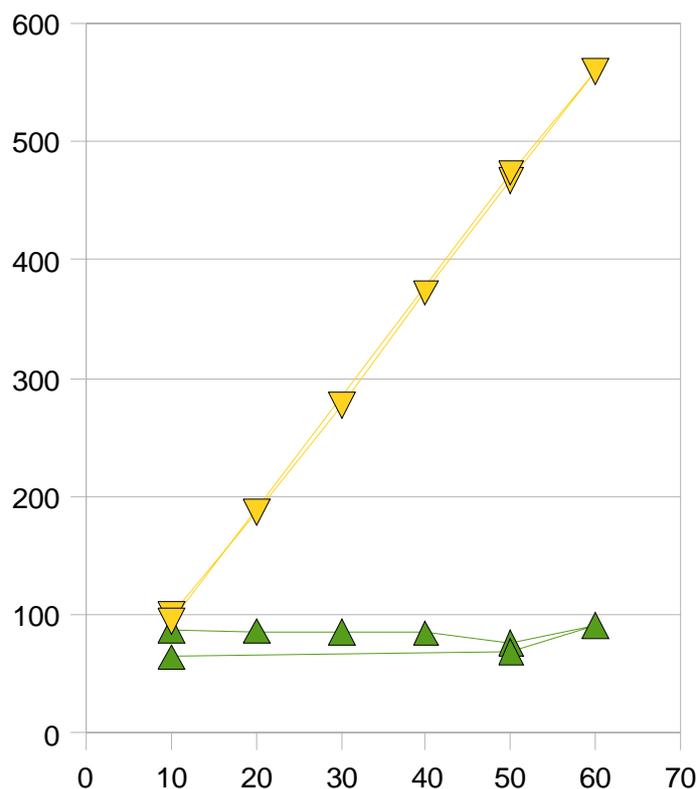

Рис.2 По оси абсцисс — напряжение источника. Треугольник — точки соответствующие току через образец, перевернутый треугольник — напряжение непосредственно на пленке в условных единицах

Однако наиболее интересное поведение электросопротивления ПВХ пленки, в литературе, по-видимому, ранее не описанное, состоит в возникновении у пластифицированной модификатором типа «А» ПВХ пленки второго устойчивого состояния с удельным сопротивлением на три порядка ниже, чем у исходной «свежей» только что изготовленная пленки. Отметим, что ранее скачкообразное изменение сопротивления тонкой пленки ПВХ ( и в десятке других полимерах) наблюдалось при воздействии высокого давления [4]. При этом «перескоки» во второе устойчивое состояние и обратно происходят случайным образом как в результате небольших изменений приложенного внешнего напряжения, так и спонтанно, при фиксированном напряжении

На рис. 3 приведен пример измерения вольтамперной характеристики образца аналогичный рис. 2, откуда видно, что в моменты переключения напряжения источника образец изменил сопротивление на три порядка величины. В отличие от известных [2]

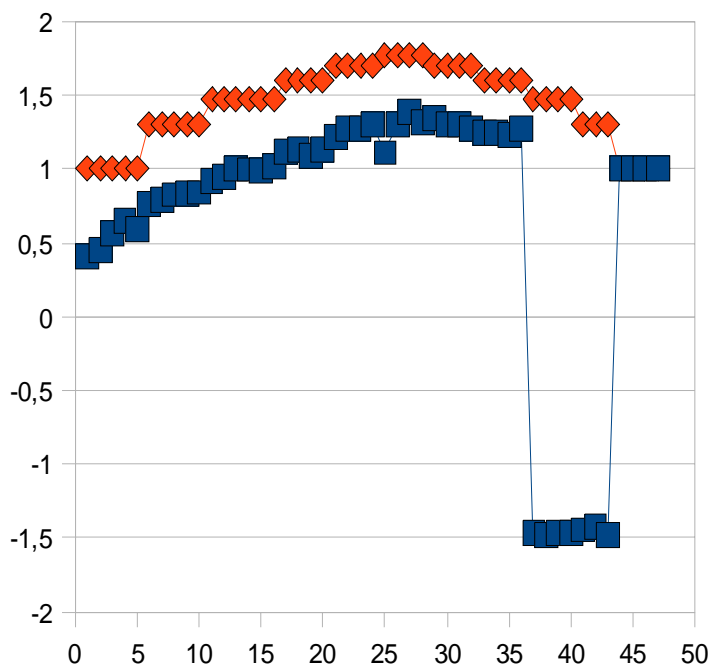

Рис.3 Зависимость сопротивления ПВХ пленки в логарифмическом масштабе при съемке вольтамперной характеристики. Прямоугольник - «сопротивление» образца, ромбики — напряжение источника. По оси абсцисс время измерения в минутах. В районе 35-45 минут — образец находился в высоко- проводящем состоянии.

случаев переключения состояний проводимости под действием электрических и магнитных полей в описываемых экспериментах переключение устойчивых состояний проводимости осуществлялись спонтанно как в процессе переключения приложенного напряжения (в частности при снижении напряжения), так и при постоянном приложенном напряжении – случай представленный на диаграмме на Рис 3.

Переход в состояние с высокой проводимостью может происходить и без переключения напряжения источника спонтанно в процессе насыщения релаксационных процессов. Состояние с высокой проводимостью является устойчивым и сохраняется при снятии напряжения иногда до нескольких суток. Независимый контроль расчетных абсолютных значений измеряемого сопротивления образца осуществлялся методом эквивалентного сопротивления, т.е. вместо образца в измерительную цепь включалось обычное сопротивление с известным номиналом. Пример независимой калибровки приведен на рис.4.

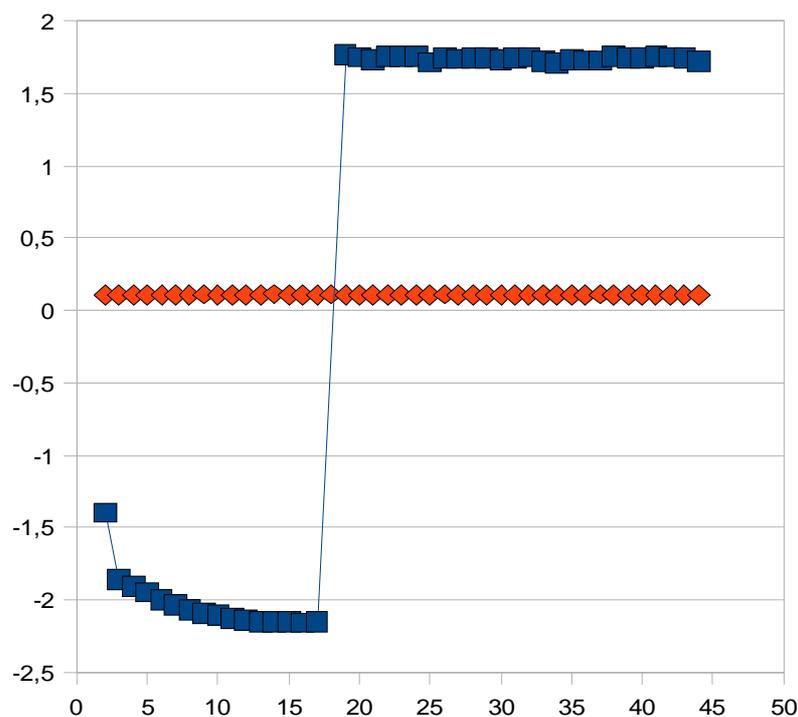

Рис.4. Пример абсолютной калибровки измерений сопротивления образца по эквивалентному сопротивлению. Зависимость логарифма отношения напряжения к току образца - Прямоугольник , прямая — зависимость логарифма отношения напряжения к току для эквивалентного сопротивления 100КОм.

Расчетные значения сопротивления образца практически совпадают с прямыми контрольными измерениями методом эквивалентного сопротивления и составляют в состоянии с высоко проводимостью -  2 КОм в состоянии с низкой проводимостью и соответственно — 2.5-11 МОм в исходном состоянии. На рис. 4 видно, что запаздывающая релаксация проводимости более явно проявляется в состоянии низкой проводимости, в то время как в состоянии высокой проводимости в пределах погрешности измерений релаксация отсутствует и соответственно значение сопротивления образца может быть определено с более высокой точностью.
   Более детальные исследования процессов переключения состояний проводимости позволят, по-видимому, прояснить физический механизм наблюдаемого переключения состояний, а также научиться управлять как процессами переключения, так и поддержания состояний заданной проводимости.

**Обсуждение результатов**.
        В работе [2] подробно обсуждаются переходы полимерных пленок в состояние с высокой проводимостью причем, приводятся результаты исследований с инициированием переходов в основном изменением давления  ( в области КБар) , электрическими и магнитными полями, легированием и температурой. В описываемых экспериментах, наблюдались спонтанные переключения состояний проводимости, причем переход в высоко- проводящее состояние мог осуществляться как при измерениях на восходящей так и на нисходящей  частей вольтамперной характеристики. Более того переход мог происходить спонтанно, как в присутствии электрического поля и соответственно тока через образец, так и в свободном состоянии, например в течение суток меду последовательными измерениями. Известно, что макромолекулярная структура ПВХ пленки может состоять из микроскопических квази-кристаллических и аморфных

образований, состоящих, в свою очередь из различных по строению молекулярных объектов (атактической, синдиотактической и изотактической природы), т.е. доменной структурой на границах которой и могут возникать обсуждавшиеся в литературе каналы проводимости ВПС.

Не углубляясь в детальный анализ механизмов переключения проводимости, которые на наш взгляд не предоставляют общей ясной и четкой картины физических механизмов переходов, отметим, что экспериментальные результаты полученные в рамках настоящей работы могут оказаться полезными с точки зрения анализа физики переключений состояний проводимости полимерных пленок. Более того, для пластифицированного модификатором «А» пленок ПВХ такие измерения выполнены впервые.

Выводы.

Разработана экспериментальная автоматизированная установка для детального анализа релаксационных и нелинейных процессов, позволяющая осуществлять длительные серии измерений вольт-амперных характеристик полимерных пленок на временах, превышающих времена установления асимптотических значений, соответствующих заданным напряжениям источника. Исследована Дебаевская запаздывающая релаксация проводимости образцов пленок при приложении напряжений существенно ниже пороговых. Измерен характер нелинейной зависимости сопротивления образцов пленок ПВХ от напряжения (эффект стабилизации тока) при приложении напряжений существенно ниже пороговых.

В результате измерений обнаружено, что при приложении напряжений существенно ниже пороговых наблюдаются спонтанные переходы между двумя относительно устойчивыми состояниями с высокой и низкой удельной проводимостью, различающиеся на три порядка. При этом каждое из состояний может сохраняться при снятии напряжения источника, в том числе и на интервалах времени более суток.

Все наблюдаемые аномалии проводимости, описанные в настоящей работе при приложении напряжений существенно ниже пороговых и полученные с применением пластификатора типа «А», снижающего сопротивление ПВХ пленки на 4 порядка по сравнению с применением стандартного пластификатора ДОФ.

Литература